\documentclass[preprint2]{aastex62}
\usepackage{graphicx}
\usepackage{xcolor}
\usepackage{epstopdf}
\usepackage{amsmath}
\usepackage{hyperref}
\usepackage{cleveref}
\usepackage{natbib}

\hypersetup{linkcolor=red,citecolor=green,filecolor=cyan,urlcolor=magenta}

\newcommand{\angstrom}{\textup{\AA}}
\def\simlt{\mathrel{\hbox{\rlap{\hbox{\lower4pt\hbox{$\sim$}}}\hbox{$<$}}}}
\def\simgt{\mathrel{\hbox{\rlap{\hbox{\lower4pt\hbox{$\sim$}}}\hbox{$>$}}}}

\usepackage[scr]{rsfso}

\graphicspath{./}

\submitjournal{ApJ}

\shorttitle{[CII]$\,158\mu m$ LFs at Redshift of 5 }
\shortauthors{Yan et al.}

\begin{document}
\title{The ALPINE-ALMA [C\,II] Survey: 

[C\,II]\,158\,micron Emission Line Luminosity Functions at $z \sim 4 - 6$}

\correspondingauthor{Lin Yan}
\email{lyan@caltech.edu}

\author[0000-0003-1710-9339]{Lin Yan}
\affil{The Caltech Optical Observatories, California Institute of Technology, Pasadena, CA 91125, USA}

\author[0000-0002-1917-1200]{A.~Sajina}
\affil{Department of Physics and Astronomy, Tufts University, Medford, MA 02155, USA}

\author{F.~Loiacono}
\affil{Dipartimento di Fisica e Astronomia, Universit\`{a} di Bologna via Gobetti 93/2, I-40129 Bologna, Italy}
\affil{Osservatorio di Astrofisica e Scienza dello Spazio di Bologna via Gobetti 93/3, I-40129 Bologna, Italy}

\author{G.~Lagache}
\affil{Aix Marseille Universit\'{e}, CNRS, CNES, LAM, Marseille, France}

\author{M.~B\'{e}thermin}
\affil{Aix Marseille Universit\'{e}, CNRS, CNES, LAM, Marseille, France}

\author{A.~Faisst}
\affil{Infrared Processing and Analysis Center, California Institute of Technology, Pasadena, CA 91125, USA} 

\author{M.~Ginolfi}
\affil{Observatoire de Gen\`{e}ve, Universit\'{e} de Gen\`{e}ve 51 Ch. des Maillettes, 1290 Versoix, Switzerland}

\author{O.~Le~F\`{e}vre}
\affil{Aix Marseille Universit\'{e}, CNRS, CNES, LAM (Laboratoire d’Astrophysique de Marseille), 13013, Marseille, France}

\author{C.~Gruppioni}
\affil{Osservatorio di Astrofisica e Scienza dello Spazio di Bologna via Gobetti 93/3, I-40129 Bologna, Italy}
\author{P.~L.~Capak}
\affil{Infrared Processing and Analysis Center, California Institute of Technology, Pasadena, CA 91125, USA} 
\author{P.~Cassata}
\affil{Dipartimento di Fisica e Astronomia, Universit\`{a} di Padova, Vicolo dell’Osservatorio, 3 35122 Padova, Italy}
\author{D.~Schaerer}
\affil{Observatoire de Gen\`{e}ve, Universit\'{e} de Gen\`{e}ve 51 Ch. des Maillettes, 1290 Versoix, Switzerland}
\affil{Institut de Recherche en Astrophysique et Plan\'{e}tologie - IRAP, CNRS, Universit\'{e} de Toulouse, UPS-OMP, 14,
avenue E. Belin, F31400 Toulouse, France}
\author{J.~D.~Silverman}
\affil{Kavli Institute for the Physics and Mathematics of the Universe, The University of Tokyo, Kashiwa, Japan 277-8583
(Kavli IPMU, WPI)}
\affil{Department of Astronomy, School of Science, The University of Tokyo, 7-3-1 Hongo, Bunkyo, Tokyo 113-0033, Japan}
\author{S.~Bardelli}
\affil{Osservatorio di Astrofisica e Scienza dello Spazio di Bologna via Gobetti 93/3, I-40129 Bologna, Italy}
\author{M.~Dessauges-Zavadsky}
\affil{Observatoire de Gen\`{e}ve, Universit\'{e} de Gen\`{e}ve 51 Ch. des Maillettes, 1290 Versoix, Switzerland}
\author{A.~Cimatti}
\affil{Dipartimento di Fisica e Astronomia, Universit\`{a} di Bologna via Gobetti 93/2, I-40129 Bologna, Italy}
\affil{Osservatorio di Astrofisica e Scienza dello Spazio di Bologna via Gobetti 93/3, I-40129 Bologna, Italy}
\author{N.~P.~Hathi}
\affil{Space Telescope Science Institute, 3700 San Martin Dr.,
Baltimore, MD 21218, USA}
\author{B.~C.~Lemaux}
\affil{Department of Physics, University of California, Davis, One Shields Ave., Davis, CA 95616, USA}
\author{E.~Ibar}
\affil{Instituto de F\'isica y Astronom\'ia, Universidad de Valpara\'iso, Avda. Gran Breta\~na 1111, Valpara\'iso, Chile}
\author{G.~C.~Jones}
\affil{Cavendish Laboratory, University of Cambridge, 19 J. J. Thomson
Ave., Cambridge CB3 0HE, UK}
\affil{Kavli Institute for Cosmology, University of Cambridge,
Madingley Road, Cambridge CB3 0HA, UK}
\author[0000-0002-6610-2048]{Anton M. Koekemoer}
\affiliation{Space Telescope Science Institute, 3700 San Martin Dr.,
Baltimore, MD 21218, USA}
\author{P.~A.~Oesch}
\affil{Observatoire de Gen\`{e}ve, Universit\'{e} de Gen\`{e}ve 51 Ch. des Maillettes, 1290 Versoix, Switzerland}
\affil{Cosmic Dawn Center (DAWN)}
\author{M.~Talia}
\affil{Dipartimento di Fisica e Astronomia, Universit\`{a} di Bologna via Gobetti 93/2, I-40129 Bologna, Italy}
\affil{Osservatorio di Astrofisica e Scienza dello Spazio di Bologna via Gobetti 93/3, I-40129 Bologna, Italy}
\author{F~.Pozzi}
\affil{Dipartimento di Fisica e Astronomia, Universit\`{a} di Bologna via Gobetti 93/2, I-40129 Bologna, Italy}
\affil{Osservatorio di Astrofisica e Scienza dello Spazio di Bologna via Gobetti 93/3, I-40129 Bologna, Italy}
\author{D.~A.~Riechers}
\affil{Cornell University, Space Sciences Building, Ithaca, NY 14853, USA}
\author{L.~A.~M.~Tasca}
\affil{Aix Marseille Universit\'{e}, CNRS, CNES, LAM (Laboratoire d’Astrophysique de Marseille), 13013, Marseille, France}
\author{Sune Toft}
\affil{Cosmic Dawn Center (DAWN)}
\affil{Niels Bohr Institute, University of Copenhagen, Lyngbyvej 2, DK-2100 Copenhagen, Denmark}
\author{L.~Vallini}
\affil{Leiden Observatory, Leiden University, PO Box 9500, NL-2300 RA Leiden, the Netherlands}
\author{D.~Vergani}
\affil{Osservatorio di Astrofisica e Scienza dello Spazio di Bologna via Gobetti 93/3, I-40129 Bologna, Italy}
\author{G.~Zamorani}
\affil{Osservatorio di Astrofisica e Scienza dello Spazio di Bologna via Gobetti 93/3, I-40129 Bologna, Italy}
\author{E.~Zucca }
\affil{Osservatorio di Astrofisica e Scienza dello Spazio di Bologna via Gobetti 93/3, I-40129 Bologna, Italy}

\begin{abstract}
We present the [C\,II]\,158$\mu$m line luminosity functions (LFs) at $z\sim4-6$ using the ALMA observations of 118 sources, which are selected to have UV luminosity $M_{1500\angstrom}<-20.2$ and optical spectroscopic redshifts in COSMOS and ECDF-S. Of the 118 targets, 75 have significant [C\,II] detections and 43 are upper limits. This is by far the largest sample of [C\,II] detections which allows us to set constraints to the volume density of [C\,II] emitters at $z\sim4-6$. But because this is a UV-selected sample, we are missing [C\,II]-bright but UV-faint sources making our constraints strict lower limits. Our derived LFs are statistically consistent with the $z\sim0$ [C\,II] LF at $10^{8.25} - 10^{9.75}L_\odot$. We compare our results with the upper limits of the [C\,II] LF derived from serendipitous sources in the ALPINE maps (Loiacono et al. 2020). We also infer the [C\,II] LFs based on published far-IR and CO LFs at $z\sim4-6$.
Combining our robust lower limits with these additional estimates, we set further constraints to the true number density of [C\,II] emitters at $z\sim 4 - 6$. These additional LF estimates are 
largely above our LF at $L_{[CII]}>10^9L_{\odot}$, suggesting that UV-faint but [C\,II]-bright sources likely make a significant contributions to the [C\,II] emitter volume density. When we include all the LF estimates, we find that available model predictions underestimate the number densities of [C\,II] emitters at $z\sim4-6$. Finally, we set a constraint on the molecular gas mass density at $z\sim4-6$, with $\rho_{mol} \sim (2-7)\times10^7M_\odot$\,Mpc$^{-3}$. This is broadly consistent with previous studies. 

\end{abstract}


\section{Introduction}
\label{sec:intr}

Galaxies in the distant Universe are thought to be more gas rich than low redshift systems. CO observations using Plateau de Bure Interferometer (PdBI) and NOrthern Extended Millimeter Array (NOEMA) have provided quantitative measurements of gas fractions at $z\sim 1 - 2$ \citep{Freundlich2019}. However, at high redshifts, studies of gas content of galaxies have not been possible for large number of sources until the advent of the Atacama Large Millimeter/submillimeter Array (ALMA). This amazing facility has opened up new vistas for astronomical research, particularly for the high redshift Universe. In particular, it allows us to probe the InterStellar Medium (ISM) of galaxies in the early Universe, through far-infrared fine structure lines. It is now possible to measure both dust content using far-infrared/submillimeter continuum emission as well as ionized and molecular gas using  far-infrared fine structure lines from galaxies in the early Universe.

The [C\,II]\,158$\mu$m emission is the strongest far-infrared fine-structure line. It can reach as high as a few percent of the galaxy total infrared luminosity. The [C\,II] line arises from collisionally excited C$^+$ ions. The ionization potential of C$^+$ is quite shallow, only 11.26\,eV, and the critical electron number density is also small, $\sim 3\times 10^3 $\,cm$^{-3}$ at $T = 100$\,K 
\citep{Goldsmith2012}. Both of these factors make the [C\,II] line the most efficient and dominant coolant for a variety of ISM, including neutral and ionized diffuse ISM.

Because of its high luminosity, the [C\,II]\,158$\mu$m line is considered the best probe of gas content in galaxies at redshift $>4$ when this transition moves into the ALMA band 7. Since ALMA operations began in 2011, many individual detections of [C\,II]\,158$\mu$m lines at $z>4$ have been published for a variety of objects, including normal star forming galaxies, QSOs and Ultra-luminous infrared galaxies (ULIRG) \citep[e.g.][]{Wang2013,Riechers2014,Willott2015,Capak2015,Emonts2015,Barro2017,Fudamoto2017,Zanella2018}. In particular, the superb spatial resolution from ALMA has revealed many stunning discoveries of large, extended gas reservoir, gaseous companions and the associated kinematics \citep{Jones2017,Shao2017,Lelli2018}.

Today, the ever increasing number of [C\,II]\,158$\mu$m detections at high redshift from ALMA has naturally led to several questions regarding their population statistics. What is the volume density of [C\,II] emitters at $z \sim 4 - 6$? What is the [C\,II] line luminosity function (LF) at $z\sim 4 - 6$? Does it evolve strongly with redshift in comparison with that of low-$z$?  The recently published [C\,II] LF at $z\sim 0$ \citep{Hemmati2017} serves as the useful local bench-mark when we discuss the redshift evolution. 

In recent years, quite a few theoretical studies were published to model the observed CO and [C\,II] emission at high redshifts \citep{Popping2016, Vallini2015, Katz2017, Olsen2017, Carniani2018, Lagache2018, Pallottini2019, Ferrara2019}. 
Most of these cosmological simulations including radiative transfer are focused on understanding the correlation and deviations around the [C\,II] luminosity and star formation rate relation. Although some of these models can reproduce $z\sim 4- 5$ luminous [C\,II] emitters found by the ALMA observations of \citet{Capak2015} and \citet{Willott2015}, only two studies made predictions on [C\,II] LFs \citep{Popping2016, Lagache2018}. However, our measurements of [C\,II] LF at $z\sim0$ \citep{Hemmati2017} has already ruled out the \citet{Popping2016} model. Because [C\,II] emission involves many complex processes \citep{Ferrara2019}, it is now becoming urgent  
to properly measure the [C\,II] line LF at $z\sim 4 - 6$ and provide constraints to theoretical models.

The goal of this paper is to address these questions, using a unique set of observations taken by the ALMA Large Program to INvestigate C\,II at Early times (ALPINE) survey \citep{lefevre2019,Faisst2020,Bethermin2020}. 
Our paper is organized as follows. \S\ref{sec:data} describes the ALPINE survey, the observations and the associated [C\,II]\,158$\mu$m line measurements. \S\ref{sec:method} discusses the methodology. The main results and conclusions are presented in \S\ref{sec:results} and \S\ref{sec:summary}.
 
12

Throughout the paper, all magnitudes are in AB system and the adopted 
cosmological parameters are $\Omega_{\Lambda,0}=0.7$, $\Omega_{M,0} = 0.3$, $\rm H_0=70$\,km\,s$^{-1}$\,Mpc$^{-1}$. 

\section{Data}
\label{sec:data}

\subsection{Summary of the ALPINE survey}
The [C\,II] emission line fluxes are measured from the ALMA data taken by ALPINE. A detailed description of this survey is discussed in \citet{lefevre2019}. The reduction of ALPINE data and the association with the extensive multi-wavelength ancillary data are presented in B\'{e}thermin et al. (2020) and \citet{Faisst2019}, respectively. A summary of the salient points of this survey is following.

\begin{figure}[!h]
    \includegraphics[scale=0.45]{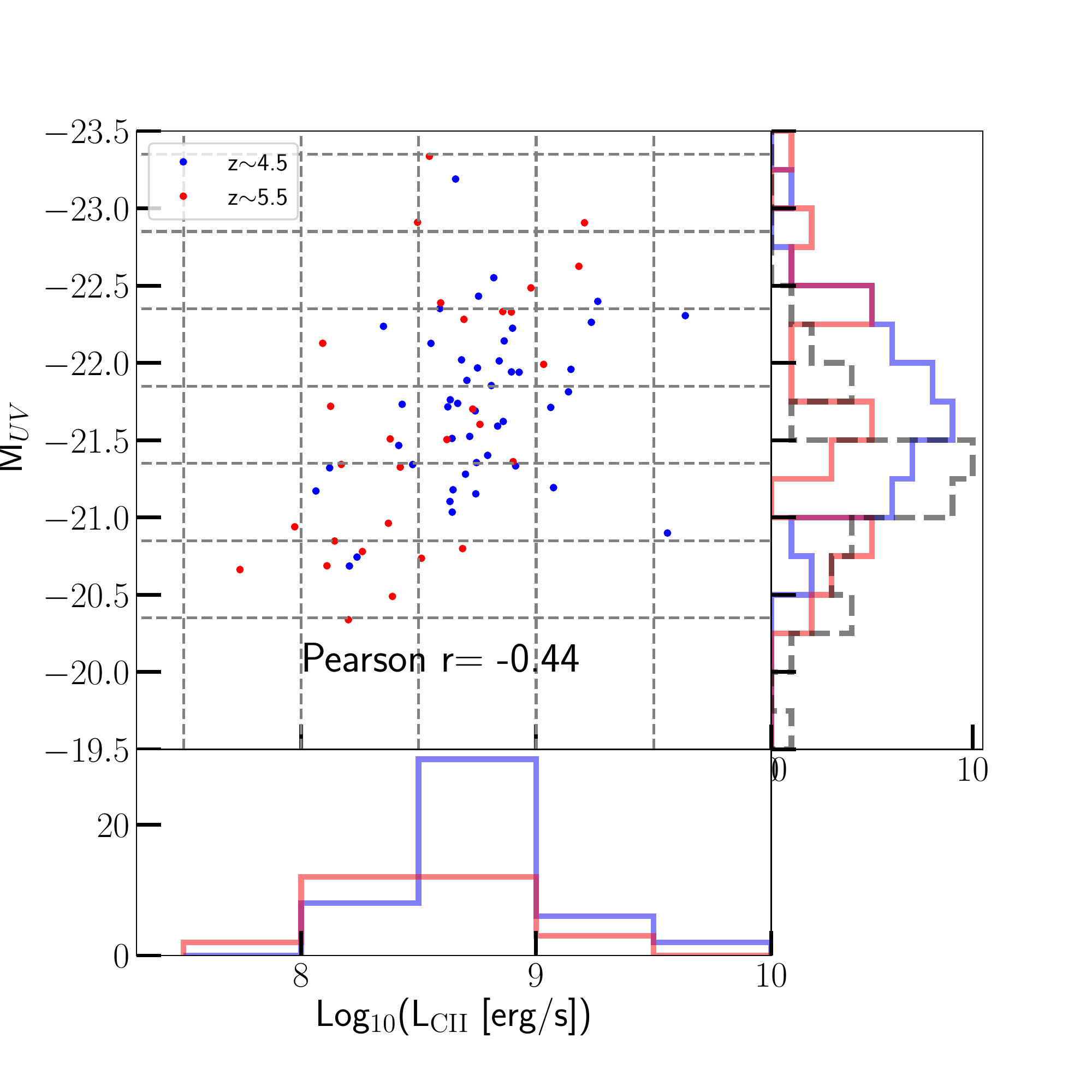}
    \caption{The [C\,II] luminosities vs observed UV magnitudes for the ALPINE sample. The Pearson $r$ coefficient indicates a weak correlation between these quantities, but with significant scatter. The  right and bottom panels show  the $M_{UV}$ and $L_{[CII]}$ distributions  for  the  sub-samples  with  [C\,II]  detections,  whereas  the  open  dashed  histogram  shows the  same  distribution  for  the  [C\,II]  non-detected ALPINE sources.  We note that the $M_{UV}$ of three [C\,II] non-detections are now below the initial cut-off of $-20.2$ due to the photometry updates \citep{Faisst2020}.
     }
    \label{fig:muv_lcii}
\end{figure}

ALPINE is a targeted survey, obtaining the [C\,II] spectral line and continuum observations with ALMA band 7 (275 - 373\,GHz) for 118 main sequence galaxies, selected by their UV luminosity at 1500\,\AA\ (Note: not by stellar mass). These 118 galaxies are chosen to have optical spectroscopic redshifts in two intervals, $z \sim 4.40 - 4.58$ and $z \sim 5.14 - 5.85$ and to have 1500\,\AA\ absolute magnitude $M_{1500\angstrom} \leq -20.2$\,mag. These targets have stellar masses of $10^{8.5} - 10^{11}M_\odot$ with an average value of $10^{9.7}M_\odot$ \citep{Faisst2019}. The redshift slice of $4.6 - 5.12$ was excluded because the redshifted [C\,II] lines would fall in a low transmission window for ALMA. The ALPINE primary targets are relatively luminous, almost all having $L_{1500\angstrom} > 0.5L^*$, corresponding to $M_{1500\angstrom} < -20.2$. With the updated photometric measurements, three [C\,II] non-detections now have slightly fainter $M_{1500}$ magnitudes, as shown by the dashed lines in Figure~\ref{fig:muv_lcii}.  For comparison, the UV LF at $z\simeq 4.5 - 5.5$ has $M^*_{1500\angstrom}=-20.96$  \citep{Ono2018}. The sample median value of $L_{1500}$ is 1.5$L^*$.  

\subsection{The [C\,II] emission line measurements}
\label{sec:line}
As discussed in B{\'{e}}thermin et al. (2020), the mean [C\,II] line flux root mean square (RMS) is 0.14\,Jy\,km/s for this sample. The spectral maps have yielded 75 [C\,II] emission lines with the peak flux to noise ratio $\rm SNR_{peak}\geq3.5\sigma$ and 43 non-detections ($<3.5\sigma$). In contrast to the high fraction of [C\,II] detections, the continuum maps have found only
23 detections at $\geq3\sigma$. The high fraction of [C\,II] detection supports the correlation between the integrated [C\,II] luminosity and star formation rate (SFR). Schaerer et al. (2020) investigates this topic using the ALPINE data. 

Another significant result is that a good fraction of the $z\sim 4 - 6$ sources is resolved -- both spatially and kinematically -- and many of them have multiple components and are clearly interacting.  Several ALPINE papers (Le F{\`{e}}vre et al. 2019; B\'{e}thermin et al. 2020; Loiacono et al. 2020) discuss this property. \citet{Jones2019} has focused on one triple merger system, deimos\_cosmos\_818760, with three [C\,II] components. Therefore, measuring the total line flux correctly is important for our calculations. 

Since a large fraction of our sources are marginally resolved, their total line fluxes are generally larger than their peak fluxes, which are good representations of the total line fluxes only for point sources. \citet{Bethermin2020} has designed three different methods to measure total line fluxes. One is a simple circular aperture photometry within a $1.5^{''}$ radius ($f_{aper}$). The second method is to sum up all fluxes within a $2\sigma$ contour above the background in the
zero-moment map ($f_{clipped}$). And the last is a 2-dimensional elliptical Gaussian fitting method ($f_{fit}$). Overall, 
these three methods give good agreement. For example, the 2D-fit and $2\sigma$-clipped fluxes are consistent, with a small offset of 3\%\ and standard deviation of only 0.17. As expected, aperture fluxes suffer larger errors, especially for faint sources (see Figure 15 in \citet{Bethermin2020}). The aperture fluxes tend to under-estimate the total fluxes compared with that of $f_{fit}$. For example, the mean and median of $f_{aper}/f_{fit}$ are 0.85 and 0.93 respectively. For this paper, we computed the LFs using $f_{fit}$ and $f_{clipped}$. As shown in Figure~\ref{fig:twomethod}, the differences are within the errors, see details in \S\ref{sec:results}.

One important and complex question is how to deal with [C\,II] sources with multiple components. 
This paper adopts the total [C\,II] flux measurements for all of the sources provided by \citet{Bethermin2020}, without deblending (see gold and red contours in Figure C.4 - 6 in that paper). The rationale is that the sample has only four systems with multiple components at comparable brightness (within a factor of a few).  
For four of these (vuds\_cosmos\_ 5100822662, deimos\_cosmos\_818760, deimos\_cosmos\_873321,
deimos\_cosmos\_434239), each has two components separated by $\leq1^{''}$ and with velocity differences $\leq200 - 300$\,km/s
(see Figure~C.1 - 6 of B\'{e}thermin et al. 2020). Our paper considers these two components are physically connected and belong to one system. Worth noting that deimos\_cosmos\_818760 is in fact a triple system, with a third, fainter component $\ge 2^{''}$ away. Our calculation does not include the fluxes from this third system (see \citet{Jones2019}). Finally, we comment on vuds\_cosmos\_5101209780, whose primary target (component {\it a}) has a companion ({\it b}) at a separation of $>2^{''}$ with a flux brighter by a factor of 2. Our paper only considers Component {\it a}. Loiacono et al. (2020) regards Component {\it b} as a part of the serendipitous [C\,II] detection sample.

Most [C\,II] systems have only faint tails or slightly extended minor components. Multi-component decomposition will not produce more accurate measurement of LF because, as we discuss in the sections below, our derived LF is computed in luminosity bin width of 0.5\,dex. Small flux boosting due to faint companions will not change our results. 

\section{Methodology}
\subsection{Luminosity function calculation}
\label{sec:method}
Our [C\,II] sample is selected by a combination of UV luminosity, redshift and [C\,II] line flux cuts. The redshift cuts produce a volume limited sample. The third selection imposes ALMA [C\,II] flux limits which vary from pointing to pointing. Such a unique selection makes the LF calculation in some ways simpler but can also be confusing. 

We first describe the basic concept of our method. 
Figure~\ref{fig:muv_lcii} displays UV absolute magnitude $M_{UV}$ versus [C\,II] luminosity $\rm log_{10}(L_{[CII]})$ for the 75 [C\,II] detections. The bottom panel of Figure~\ref{fig:muv_lcii} presents the number of [C\,II] emitters, $N_{[CII]}$, per luminosity bin $\rm \Delta log_{10}(L_{[CII]})$. In the simplest form, the [C\,II] luminosity function at the $i$th bin is 
\begin{align}
    \phi_i & = \frac{1}{\Delta (\log L_{[CII]})} \frac{N_{[CII],i}}{V_{[CII],i}}\times C_{UV,i} \label{eq:phi1} \\
        &   = \frac{1}{\Delta (\log L_{[CII]})} \frac{N_{[CII],i}}{V_{[CII],i}}\times \frac{\Delta M_{uv}\phi_{uv} V_{uv}}{N_{uv,i}}
    \label{eq:phic+}
\end{align} 
where $N_{[CII],i}$ is the number of [C\,II] emitters in the $i^{th}$ bin, $V_{[CII],i}$ is the volume available to these sources, and $C_{UV,i}$ is the incompleteness factor. This incompleteness comes from the fact that the ALPINE targets are a small sub-set of {\bf all} galaxies that meet our UV luminosity cut 
($M_{1500A}<-20.2$) 
among the parent galaxy population. As shown in Equation~\ref{eq:phic+}, $C_{UV,i}$ is equal to the ratio of the number of expected UV galaxies from the known UV LF to the number of ALPINE UV galaxies within a volume element $V_{uv}$. Note that if the $\log L_{[CII]}$ to $M_{UV}$ relation were exactly 1:1, the above could be directly computed as written for each luminosity bin. However, Figure\,\ref{fig:muv_lcii} shows the two variables have significant scatter. This means that in fact we need effectively a weighted $C_{UV}$ that is based on the $M_{UV}$ distribution among all ALPINE sources within the $i^{th}$ [CII] luminosity bin or in other words use $C_{UV,i}=(1/N_{[CII],i})\sum_{j=1}^{N_{[CII],i}} C_{UV,j}$ where $j$ is the index across all $M_{UV}$ bins sampled by the ALPINE sources. 

The second issue is that the ALPINE survey is a targetted survey as opposed to a flux-limited survey which means one would expect the volume elements $V_{[CII]}$ and $V_{UV}$ to exactly match and therefore cancel out in Equation\,\ref{eq:phic+} (where the two volumes appear in respectively in the denominator and nominator). Recall that a co-moving volume is computed as $A\times [D_M^3(z_{lower}) - D_M^3(z_{upper})]$, where $A$ is the areal coverage and $D_M$ is the co-moving distance, a function of redshift. Now, whatever $A$ we choose to adopt in our calculation of $V_{[CII]}$ has to be the same as the one used to calculate the expected number of UV targets in our survey. Thus, the areas in the denominator and nominator cancel. {\it For the ALPINE survey, the [C\,II] LF calculation does not rely on the areal coverage value} and we only need to know the values of $z_{lower}$ and $z_{upper}$. For the UV galaxies, by selection, the volume is defined by its redshift slices with $z_{lower}$ and $z_{upper}$ set as $4.403 - 4.585$ and $5.135 - 5.85$ respectively for the two slices. However, for the [C\,II] detections, the volume is {\bf not} the same as defined by the redshift slices because each ALMA pointing has different sensitivity limits. For all [C\,II] sources in the $z\sim4.5$ bin, their ALMA maps are deep enough, so that their maximum allowed redshifts (thus volumes) are the same as the upper limits of the redshift slice ($z_{upper}$). At $z\sim5.5$ bin, 8 of 29 sources have slightly shallower ALMA maps, therefore, their maximum allowed (or maximum detectable) redshifts are slightly less than $z_{upper} = 5.85$. Therefore, the ALPINE survey does have a flux limited element. So for the vast majority of sources, the volume elements in Equation\,\ref{eq:phic+} exactly cancel. But for 8 of our source this is not the case. Putting these arguments together, we have a final version of our LF calculation per luminosity bin $i$, given by Equation\,\ref{eq:totalphi}.   
\begin{equation} \label{eq:totalphi}
\phi_i = \frac{1}{\Delta (\log L_{[CII]})} \sum_{j=1}^{N_i} \frac{C_{UV,j}}{V_{[CII],j}} \end{equation}
where $\phi$ is for the $i$th luminosity bin, and summation goes through all [C\,II] sources within this bin with $j$=1,2,...$N_i$. The reader will note that this equation is just a modified $1/V_{max}$ method \citep{Schmidt1968}. However, this is not the true $1/V_{max}$ method because the calculation is pegged directly to the published UV LF via the $C_{UV}$ incompleteness correction factor. Therefore any issues with this method that involve small number statistics, clustering effects, as well as knowledge of precise areal coverage are all cancelled out as long as one trusts the accuracy of the state-of-the-art UV luminosity function calculation we use \citep{Ono2018}. For example, if our targets happen to be clustered in our redshift ranges, leading to number density of [CII] emitters enhanced by some factor, the ratio of expected UV sources to known UV sources within the ALPINE target would be slightly lower by the came factor fully cancelling it out.

\begin{figure}[!h]
    \includegraphics[scale=0.45]{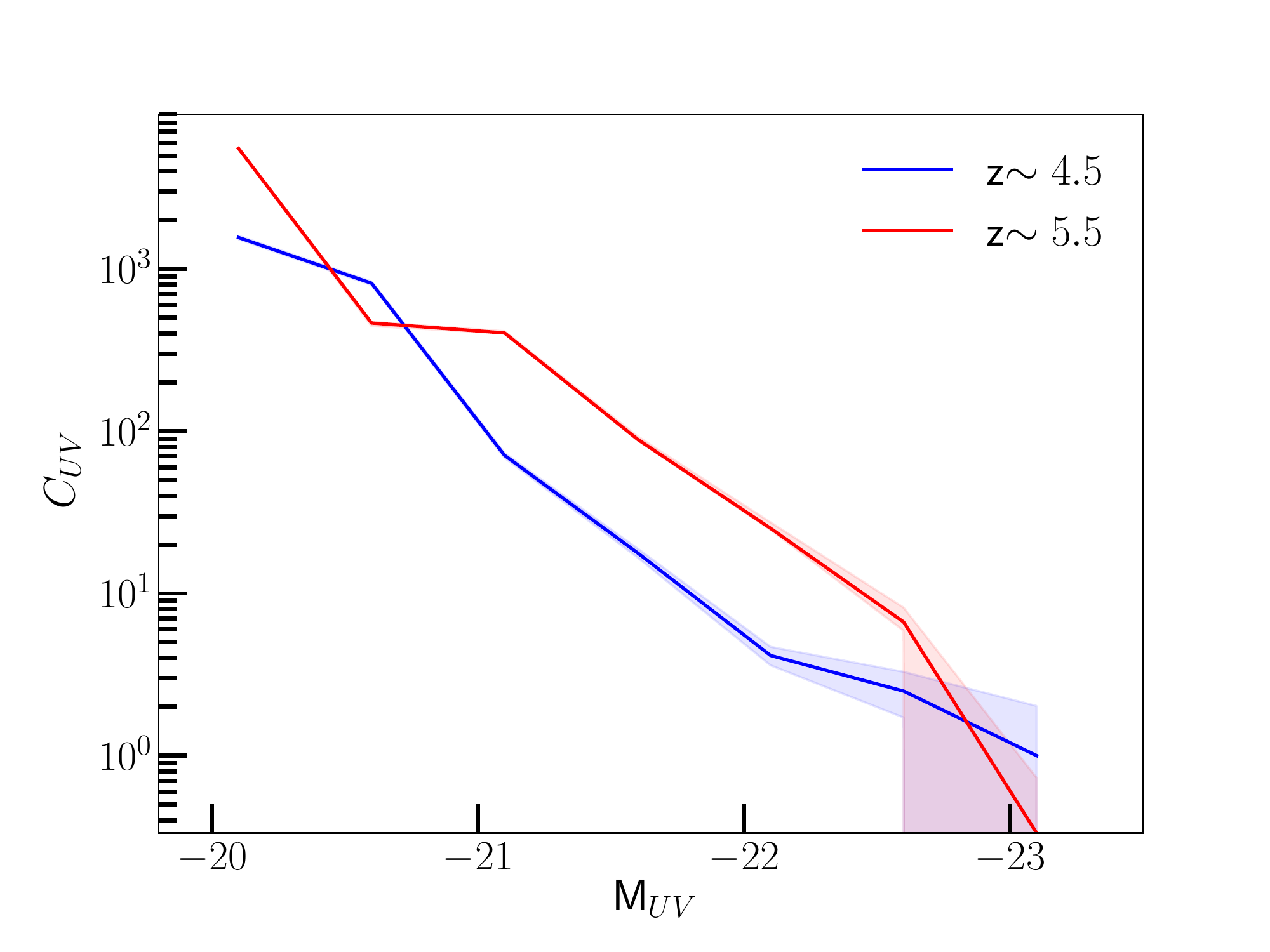}
    \caption{The incompleteness correction $C_{UV}$ as a function of absolute UV magnitude for the two redshift bins. The shaded regions are uncertainties/errors on $C_{UV}$  computed using Monte Carlo simulations as discussed in Section\,\ref{sec:method}.}
    \label{fig:correction}
\end{figure}

Figure\,\ref{fig:correction} shows our calculated $C_{UV}$ for both the two redshift bins.  For the sake of this calculation and display, we adopted $A=2$\,sq.deg (size of the COSMOS field) but this assumption cancels out in the LF calcualtion as discussed above. Here $C_{UV}$ was calculated in bins of UV magnitudes of width 0.5, but these values were interpolated so that we can have $C_{UV,j}$ corrections corresponding to each individual source's UV magnitudes in Equation\,\ref{eq:totalphi}. In Figure\,\ref{fig:correction}, the error on $C_{UV}$ is derived from the standard deviation of the expected number of galaxies per UV magnitude bin calculated from 1000 Poisson draws per bin. This Poisson uncertainty dominates the uncertainty of the UV LF \citep{Ono2018}.  

\subsubsection{Error estimates for the [C\,II] LF}
In Equation\,\ref{eq:phi1}, we see that our estimates of the LF per [C\,II] luminosity bin are based on the product of $N_{[C\,II],i}$ and the UV-incompleteness correction factor $C_{UV,i}$. As such the uncertainty in $\phi_i$ is given by the usual rule of error propagation for products:
\begin{equation}
    \sigma(\phi_i)=\phi_i\sqrt{\Big(\frac{\delta_{N_{[CII],i}}}{N_{[CII],i}}\Big)^2+\Big(\frac{\delta_{C_{UV,i}}}{C_{UV,i}}\Big)^2}
    \label{eqn:phi_error}
\end{equation}

The first of the error terms, the fractional error on the number of sources per [CII] bin is derived assuming Poisson statistics and based on the lower and upper 1\,$\sigma$ uncertainties tabulated in \citet{Gehrels1986}. Due to the relatively small number of sources in all of our bins, this turns out to be the dominant error term. 
The second term uses the $C_{UV}$ uncertainties as described above and shown in Figure\,\ref{fig:correction}. Recall that $C_{UV,i}=(1/N_{[CII],i})\sum_{j=1}^{N_{[CII],i}} C_{UV,j}$ which translates to $\frac{\delta_{C_{UV,i}}}{C_{UV,i}}=\frac{\sqrt{\sum \delta^2_{C_{UV,j}}}}{\sum C_{UV,j}}$. 

\subsection{Accounting for [C\,II] non-detections}
In our method above, we already accounted for the incompleteness of the ALPINE sample relative to its parent sample of $M_{UV}<-20.2$ sources. However, there is a second type of incompleteness due to [C\,II] non-detections.
Of the 118 ALPINE targets, 43 sources only have [C\,II] flux limits, with 22 and 21 objects in the $z\sim4.5$ and the $z\sim5.5$ (optical spectroscopic redshifts) bin respectively. To account for their maximum potential contribution to the [C\,II] LFs, we take the non-detections noise estimates, $\sigma$, and assign them a maximum luminosity of 3$\sigma$. This is a reasonable upper limit for the non-detections, since the distribution of SNR$_{peak}$ for non-detections peaks at $\sim2$ (B\'{e}thermin et al. 2020) and the threshold for detection is SNR$_{peak} \sim 3.5$.

\begin{figure}[h!]
    \includegraphics[scale=0.38]{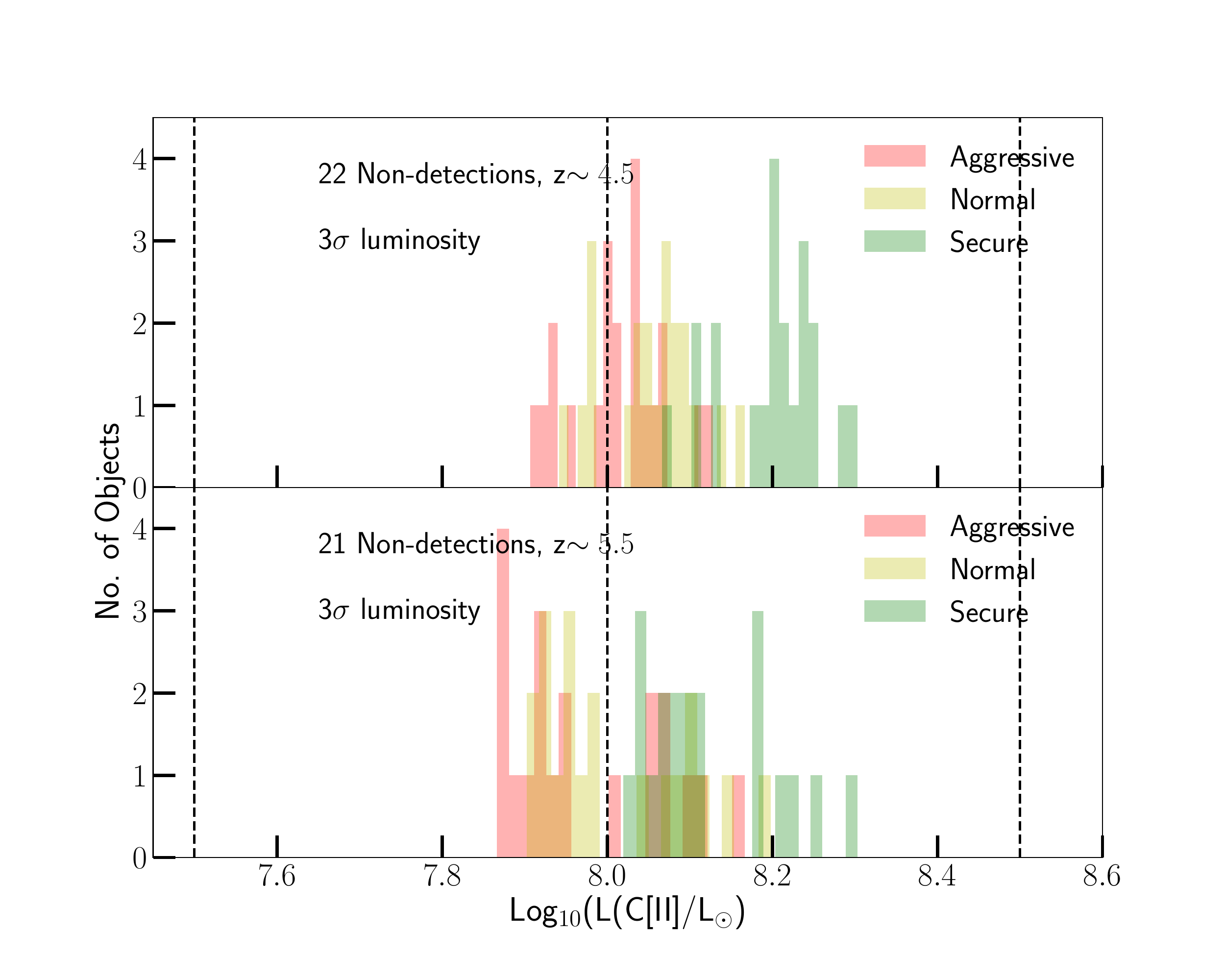}
    \caption{The distributions of $3\sigma$ luminosity of non-detected [C\,II] sources, measured using three different methods. The dashed lines mark the two lowest luminosity bins used for the [C\,II] LF calculations. \label{fig:noise}}
\end{figure}

Figure\,\ref{fig:noise} shows the contribution of non-detections to our two lowest luminosity bins. We consider three different measurements of the noise, aggressive (lowest noise), normal and secure (highest noise) $\sigma$, which are described in Section 3.3 in \citet{Bethermin2020}. Figure~\ref{fig:noise} shows 
that the distribution of $3\sigma$ upper limits measured by the secure method is systematically shifted to higher luminosity roughly by 0.2\,dex compared to that by the aggressive method. We consider the maximum potential contribution of non-detections to each luminosity bin to come from whichever noise estimate maximizes the number of non-detections in the particular bin (i.e. aggressive method at the lower luminosity bin, and the secure method in the higher luminosity bin). Because the non-detections could in principle lie below the 3\,$\sigma$ limit, these are the upper limits. We remind the reader that galaxies that are treated with such upper limits lie exclusively in the lowest two luminosity bins so in all other bins, as we discuss below, our estimates are in fact strict lower limits. In forthcoming figures we show both the LFs computed from the detected sources alone, and those that include the maximum potential contribution from the non-detections.  

The last type of incompleteness is that ALPINE is not a blind [C\,II] emission line survey. Rather it is a survey specifically targeting galaxies selected by 1500\AA\ UV luminosity. The high fraction of [C\,II] detections in the ALPINE sample implies that the UV selection should detect [C\,II] emitters efficiently, however the UV luminosity and [C\,II] line luminosity do not have a tight linear correlation (see Figure\,\ref{fig:muv_lcii}). The ALPINE selection would certainly miss luminous [C\,II] emitters with low UV luminosity, {\it i.e.} $M_{1500\angstrom} > -20.2$\,magnitude. {\it Thus, the calculated LF based on the UV-selected sample sets the lower bounds to the [C\,II] LF.} Including our non-detections by assuming they are at 3\,$\sigma$ pushes our estimates up, in particular in our two lowest luminosity bins. However, these are still contributions from sources above our UV magnitude limit. To fully account for [C\,II]-emitters that are potentially UV-faint -- i.e. below the ALPINE target selection threshold, we need a blind [C\,II] survey. The ALPINE surveys allows for the next best thing -- i.e. search for serendipitous sources within the same ALMA data as our targets. This search was conducted by Loiacono et al. (2020).  In Section~\ref{sec:resultsser} we compare our results with those of Loiacono et al. (2020) who compute the [C\,II] LFs using serendipitous [C\,II] sources in the ALPINE survey. Because of potentially uncertain redshift identifications and strong clustering effects, the results from the serendipitous sources should be considered upper bounds. The comparison of the two estimates of the $z\sim4-6$ [C\,II] luminosity function therefore helps to overcome some of the systematic uncertainties associated with each method and affirm the overall results on the number density and luminosity function of [C\,II] emitters at these redshifts as seen from the ALPINE survey.

\section{Results and discussion}
\label{sec:results}
\subsection{Measurements based on the ALPINE UV sample \label{sec:results_uvsample}}

\begin{figure}[!h]
    \includegraphics[scale=0.43]{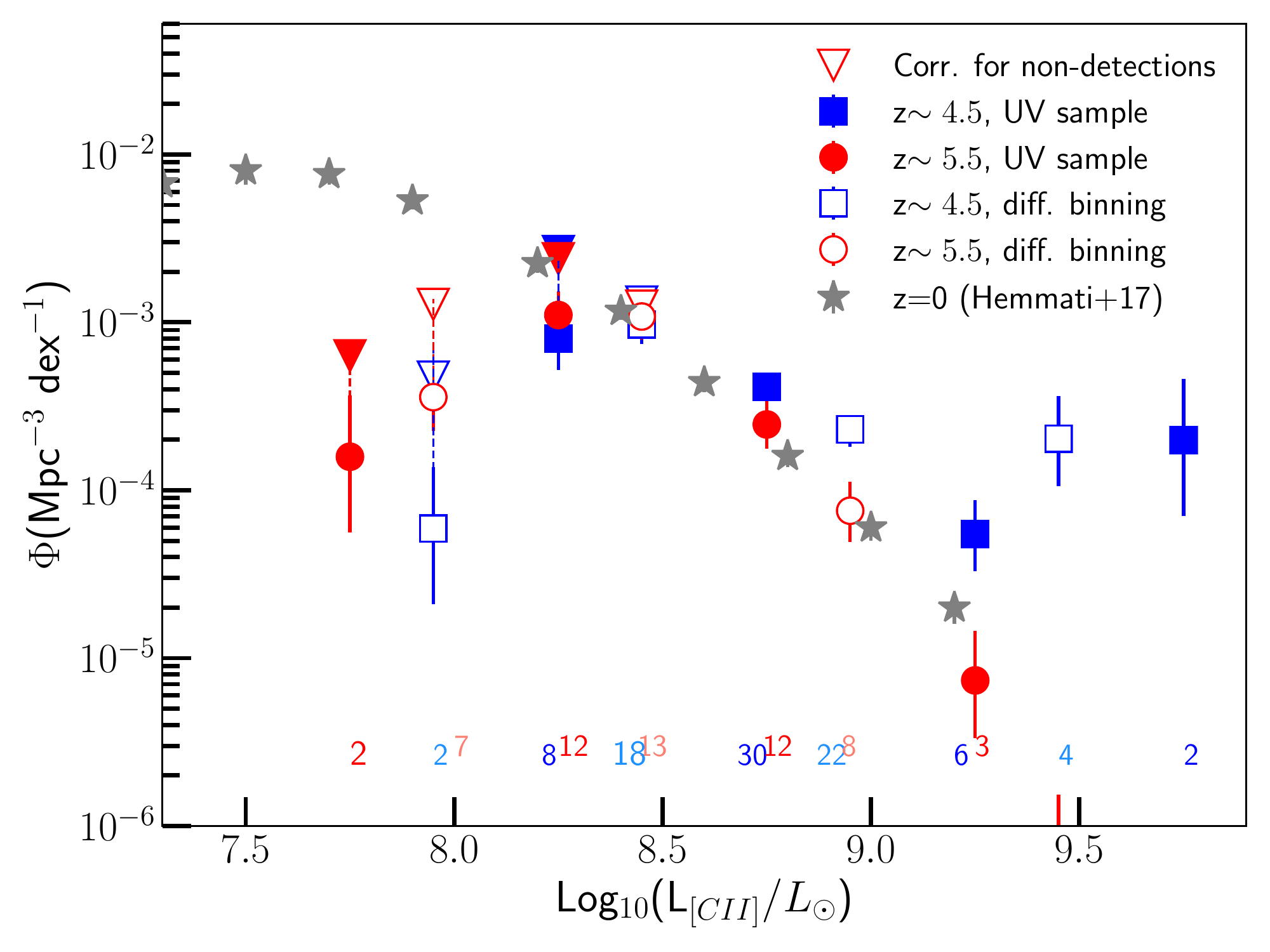}
    \caption{The robust lower bounds on the [C\,II] emission line luminosity functions at redshift $\sim4.5$ (blue squares) and $\sim5.5$ (red dots).   The open symbols of the same colors use a different binning than the solid points.
    The gray stars are the [C\,II] LF at $z\sim0$ \citep{Hemmati2017}. The down-ward triangles indicate the upper limits estimated from the [C\,II] non-detections. The numbers at the bottom indicate the number of sources in the corresponding luminosity bin and redshift. }
    \label{fig:clf}
\end{figure}

Figure~\ref{fig:clf} presents our [C\,II] LFs at $z\sim4.5$ and $z\sim5.5$, with the number of sources in each bin labeled. The down-ward arrows mark the upper limits if we include the [C\,II] non-detections at the two lowest luminosity bins. These are connected by the dashed vertical lines to the points derived using only significant [C\,II] sources. 
Table~\ref{tab:clf} gives our derived LF values at $z\sim4.5$ and $z\sim5.5$ using two different binnings.

Figure\,\ref{fig:clf} shows the following conclusions. First, the lower limits on the [C\,II] LFs at $z\sim4-6$ derived from the ALPINE UV-sample are consistent with the $z\sim0$ LFs at the luminosity range of $\rm log_{10}(L_{[CII]}/L_{\odot}) \sim 8.25 - 9.25$ (bin center, 0.5dex bin width). At $\rm L_{[CII]} \geq 10^{9.5}L_\odot$, the volume density of high-$z$ [C\,II] emitters could be higher than that of $z\sim0$, but our measurement is only at $(1-2)\sigma$ significance level (68\%\ - 95\%). Such a large uncertainty requires future confirmation. 

Second, we do not measure significant differences between [C II] LFs at $z\sim4.5$ and $z\sim5.5$, within the uncertainties. Considering that LF is essentially the volume density of [C\,II] emitters, we compute the LF at $z\sim4 - 6$ simply by using the geometric mean of the values at $z\sim4.5$ and $z\sim5.5$, except at the lowest and highest luminosity bins where only $z\sim4.5$ LF measurements are available. For these two bins, we take only the $z\sim4.5$ LF values, assuming the volume density does not change between $z\sim4.5$ and $5.5$. 
For simplicity, the subsequent figures show the [C\,II] LF in a single redshift bin of $4<z<6$.

To check the dependence of our results on the specific binning, in Figure~\ref{fig:clf} we compare our LF with that derived using a different binning  binning (open symbols). The results are consistent with each other. 

\begin{figure}
    \centering
    \includegraphics[scale=0.45]{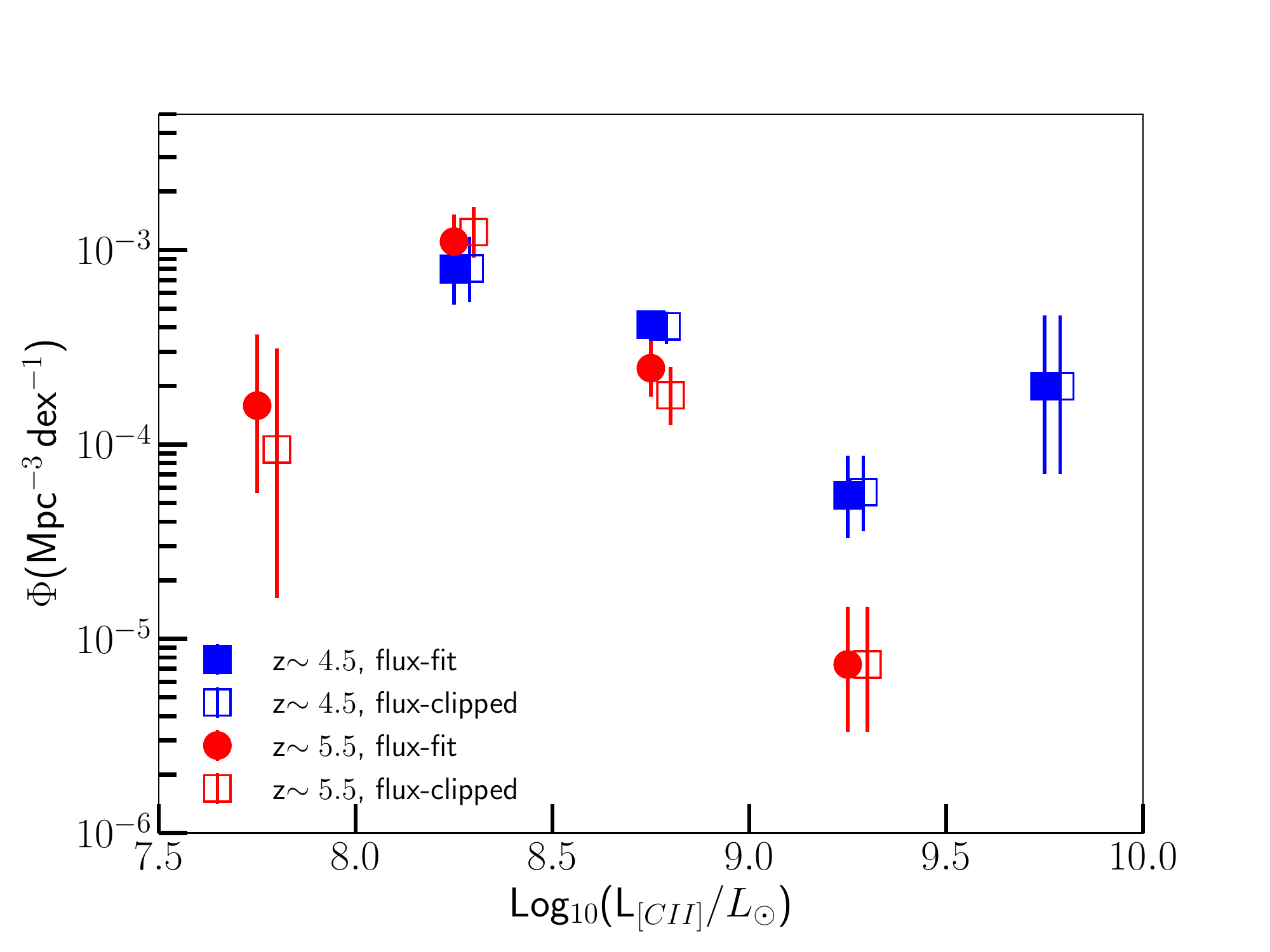}
    \caption{The [C\,II] LF computed using two different [C\,II] line fluxes, $f_{clipped}$ and $f_{fit}$}
    \label{fig:twomethod}
\end{figure}

To check the dependence of our results on the specific photometry method, Figure~\ref{fig:twomethod} compares the LFs using $f_{fit}$ (which we adopt throughout this paper) and $f_{clipped}$. Within the errors, the difference is not significant. 

To check the dependence of our results on the method for the [C\,II] luminosity function, we also examined an alternative approach. 
This method uses the $M_{UV}$ vs $L_{[CII]}$ distribution in Figure\,\ref{fig:muv_lcii} to construct the probability distribution in each UV magnitude bin for a given $L_{[C\,II]}$. The expected number density of sources from the UV LF is spread among the $L_{[C\,II]}$ bins according to these distributions. Summing the contributions from each UV bin gives us the total number density expected in each $L_{[C\,II]}$ bin. The results are fully consistent with our method described above. 

Finally, we carefully check to see if source blending can be an issue, boosting our highest lumionsity bins. We check the [C\,II] maps \citep{Bethermin2020} for the four most luminous sources included in the highest-$L$ bins in Figure~\ref{fig:clf}. 
Three of the four sources are [C\,II] bright and isolated objects, and one has an optical counterpart with very low UV luminosity. 
The fourth source, deimos\_comos\_818760, is a triple system and has been published in \citet{Jones2020}. The whole system has component C, E and W with $\rm log_{10}(L_{[CII]}/L_{\odot})=9.48$, $9.21$ and $8.7$ respectively \citep{Jones2020}. With small spatial and velocity separations, 
component C \&\ E are considered as a major merger \citep{Jones2020} and is counted as a single source with $\rm log_{10}(L_{[CII]}/L_{\odot})=9.66$ in our calculation. We remind that our LF bin width is 0.5\,dex.  Component W is included in the Loiacono et al. serendipitous sample. The UV counterpart to component C is the ALPINE primary target, however, the counterpart to component E has high dust extinction, and not detected in HST F814W band \citep{Koekemoer2007,Koekemoer2011}, but detected in deep $K$-band, with low UV luminosity, $M_{1500}$ well below the ALPINE cutoff.  Treating these two components separately will lead to two (as opposed to one) [C\,II]-bright but very UV faint sources with corresponding large $C_{UV}$ corrections and would boost our LF at the highest luminosity. We choose to treat them as one source as the more conservative approach. 
While our calculations are robust (as the above tests show), they are in fact {\it only lower bounds} on the true [C\,II] luminosity function at these redshifts. This is because any UV selected sample such as ALPINE is biased against [C\,II] emitters with low UV luminosity. 
As Figure\,\ref{fig:muv_lcii} shows, the highest luminosity bin of the [C\,II] LF indeed has one [C\,II] bright and UV-faint galaxy. Its large $C_{UV}$ noticeably boosts the LF value in this luminosity bin. This source (COSMOS\_DEIMOS\_873756) is a so-called ``4.5$\mu$m excess'' source which is off the main sequence with a large stellar mass limit ($<10^{10.2}M_\odot$) and very low SFR ($\rm Log_{10}(SFR)=0.72^{+0.5}_{-0.2}M_\odot$/yr) \citep{Faisst2020}. Note however that the stellar mass is an upper limit and the SFR has large error bars. If this source is atypical of the general UV population of the same UV magnitude, then our calculated UV incompleteness correction is not applicable here. Therefore, our estimate of the luminosity function in this highest luminosity bin has this additional systematic uncertainty that needs to be kept in mind. 

\subsection{Comparison with the 
[C\,II] LFs based on serendipitous ALPINE sources}
\label{sec:resultsser}
\begin{figure}[!h]
    \includegraphics[scale=0.45]{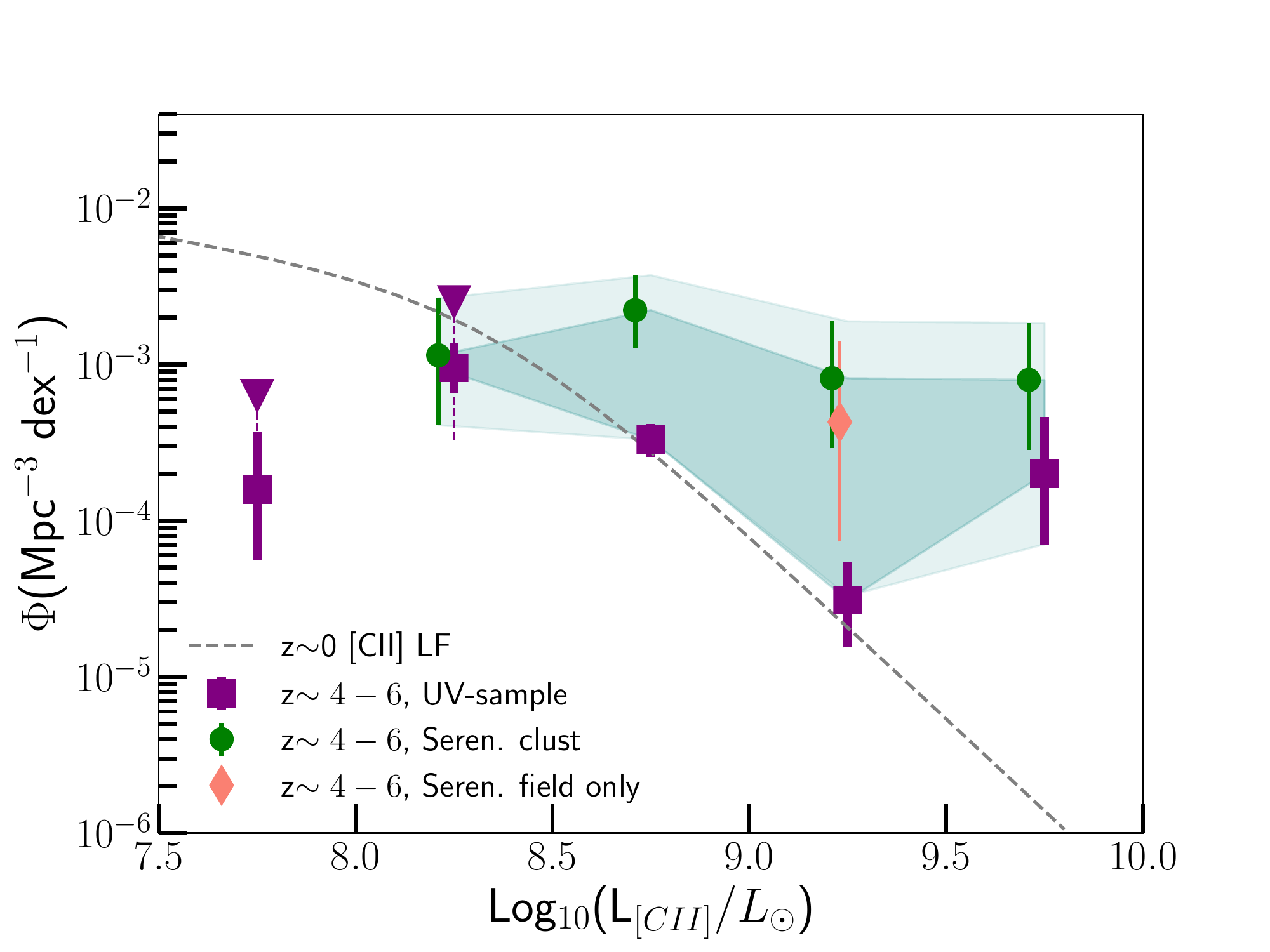}
    \caption{The comparison of [C\,II] LFs calculated from the UV sample (purple squares) with the estimates from the serendipitous [C\,II] sources which are in clusters (green) and in field (salmon). These data are taken from Loiacono et al. 2020 and the bins are independent.  The dashed line indicates the $z\sim0$ LF \citep{Hemmati2017}. The area between our LFs and that of Loiacono et al. is highlighted in green shaded region. The fainter, shaded region extends out to the upper and lower ends of the error bars. 
     }
    \label{fig:clftotal}
\end{figure}

The LFs presented above are lower limits on the number densities of [C\,II] emitters at $z\sim4-6$ because they miss potential sources that are bright in [C\,II]  but faint in UV.  Fortunately, with the 118 ALMA maps along separate pointings, searching for serendipitous [C\,II] emitters at $4.40 < z < 4.58$ and $5.13 < z< 5.85$ can quantify this effect. In a companion paper, Loiacono et al. (2020) carried out a blind search for all line emitters within the 118 data-cubes. They identified 8 [C\,II] emitters at $z\sim5$ confirmed by the optical spectroscopic redshifts, and 4 possible [C\,II] candidates with only photo-$z$. Of these 12 sources, 11 are associated with their respective primary targets in both spatial and velocity space, and 1 is truly a field object, for details, see \citet{Loiacono2020}. These 11 ``clustered''' sources are used to compute the [C\,II] LF shown as green points in Figure~\ref{fig:clftotal}. The 1 field serendipitous object is used to constraint the [C\,II] LF shown as diamond point in salmon color in Figure~\ref{fig:clftotal}.
Figure\,\ref{fig:clftotal} compares our derived LF at $z\sim 4 - 6$ with that of Loiacono et al. The LF based on the serendipitous sources is strongly affected by cosmic clustering and by uncertainties in the identification of some line emitters. Thus it sets an upper boundary to the bright-end of the [C\,II] luminosity function at $z\sim4-6$, whereas our results are the lower limits. The true [C\,II] LF lies in the shaded region, covering the purple squares and green dots and extending out to the upper and lower ends of the error bars in Figure~\ref{fig:clftotal}. 

Considering the upper and lower limits derived from the serendipitous and UV-selected ALPINE samples, we conclude that the space densities of [C\,II] emitters at $z\sim 4 - 6$ are at least as high as that of $z\sim0$. At $\rm log_{10}(L_{[C\,II]}/L_{\odot})>9.5$, there could be more high-$z$ luminous [C\,II] emitters than locally, but our estimate has large errors. This uncertain but tantalizing result should be validated by future studies.
Together, the UV-selected and serendipitous ALPINE samples both point toward significant number densities of luminous [C\,II] emitters at $z\sim4-6$. At the low luminosity end with $\rm log_{10}(L_{[CII]}/L_{\odot})\leq8.75$ both LFs are consistent with the $z\sim0$ LF, suggesting that the space density of [C\,II] emitters at $z\sim 5$ in this luminosity range is similar to that of $z\sim0$.
However, the number densities at $\rm log_{10}(L_{[C\,II]}/L_{\odot})>9$ are far in excess of the $z\sim0$ [C\,II] luminosity function of \citet{Hemmati2017} suggesting potentially considerable evolution.

\subsection{Comparison with indirect observational expectations \label{sec:results_otherobs}}

\begin{figure}[h!]
    \centering
    \includegraphics[scale=0.45]{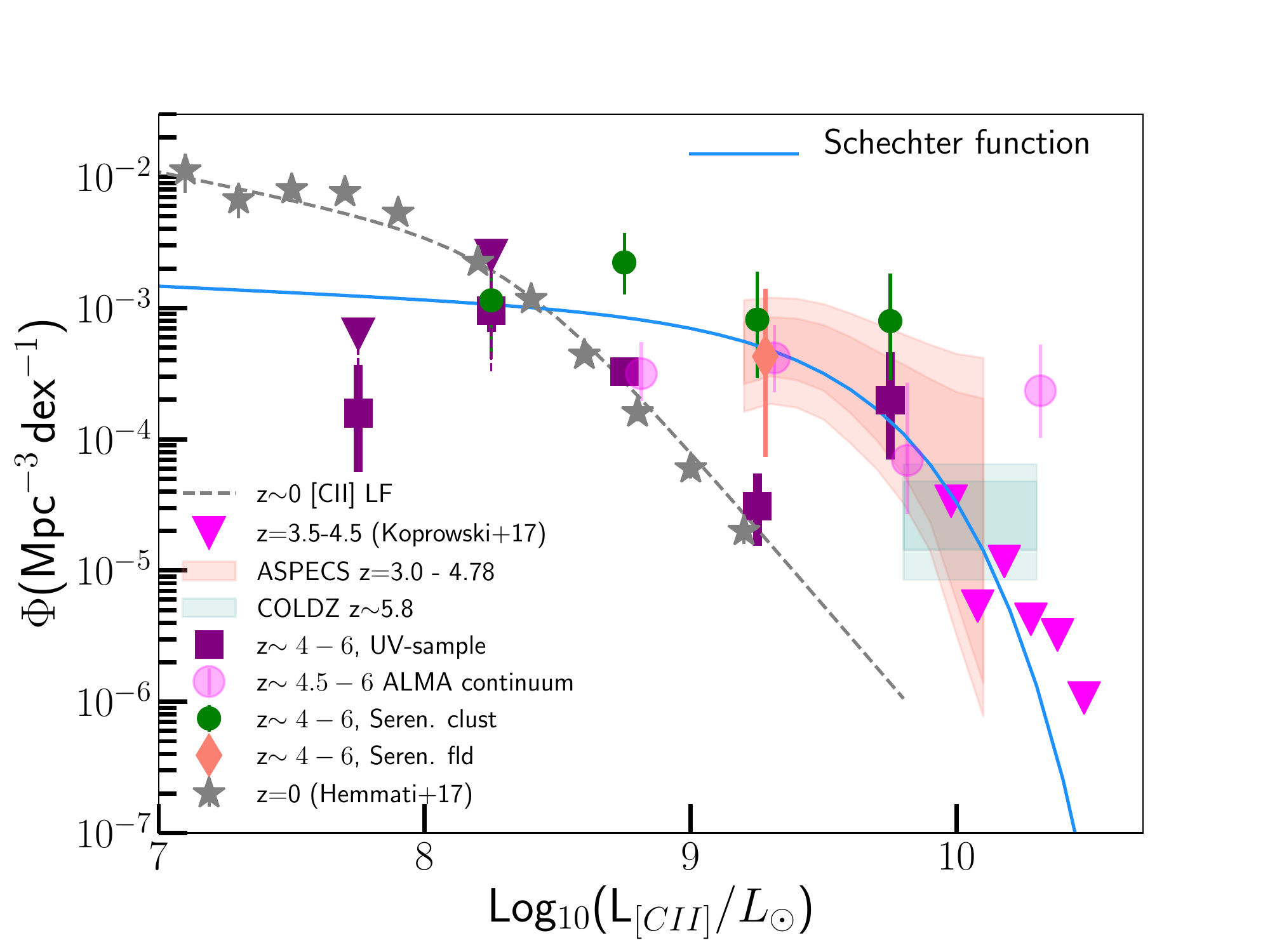}
    \caption{The [C\,II] emission line luminosity functions derived from the UV sample (purple symbols) and the serendipitous, confirmed [C\,II] emitters at $z\sim 4-6$ (green circles and red diamond). Also included are 
    the [C\,II] LFs converted from the ALPINE IR LF (light pink circles) (Gruppioni et al. 2020), from the SCUBA-2 rest-frame 250$\mu$m LF at $z\sim3.5 - 4.5$ (dark pink triangles) \citep{Koprowski2017}, and from the CO LF published by ASPECS \citep{Decarli2019} and COLDz \citep{Riechers2019}. The pink and light green shade regions represent both $1\sigma$ and $2\sigma$ uncertainty regions from \citet{Decarli2019,Riechers2019}. The blue solid line is the Schechter functional form fit to all of the data, including the estimated [C\,II] LF values based on FIR LFs
    \label{fig:clf+fir+co}}
\end{figure}

In this section, we further test our conclusion above by comparing the direct ALPINE [C\,II] LFs with indirect estimates of the [C\,II] luminosity functions based far-IR, and CO luminosity functions at $z\sim4-6$ and adopting typical conversion relations. Figure\,\ref{fig:clf+fir+co} shows this comparison. Below we describe in more detail the assumptions behind this figure. 

Our first comparison sample is again drawn from serendipitous continuum sources in the ALPINE survey. Here we use the results in Gruppioni et al. (2020) who derive an IR luminosity function based on serendipitously detected {\it continuum} sources in the ALPINE data. We also use the 250$\mu$m LF from the SCUBA-2 survey \citep{Koprowski2017}.  To convert this to a total IR luminosity function, we adopted $\rm L_{IR}/L_{250\mu m} = 38.5$ calculated using the infrared spectral energy distribution (SED) template, constructed from data including ALPINE continuum measurements by B\'{e}thermin et al. (2020). Finally, to convert both of these into [C\,II] luminosity functions 
we adopt the $\rm L_{IR}/L_{FIR} = L_{8-1000\mu m}/L_{42-122\mu m} = 1.3$ and $\rm \log_{10}(L_{FIR}/L_{[C\,II]}) = 2.69$ recently compiled for high-$z$ galaxies \citep{Zanella2018}. 

Next we include the [C\,II] LFs at $z\sim5.8$, converted from the CO(1-0) LF published by the ALMA SPECtroscopic Survey (ASPECS, pink shaded region) \citep{Decarli2019} and from the CO(2-1) LF from the CO Luminosity Density at High Redshift survey (COLDz, light shaded green) \citep{Riechers2019}. Here we use CO(2-1)/CO(1-0) ratio of 1 \citep{Riechers2019}. For comparison, the average value for this ratio is 0.85 for sub-millimeter galaxies. The ASPECS and COLDz CO LFs are published in $\rm L^{\prime}_{CO(1-0)}[K\,km\,s^{-1}\,pc^2]$.
We transformed the CO(1-0) LF to the estimated [C\,II] LF by using 
\begin{align}
& \rm \log_{10}(L_{CO(1-0)}/L_\odot) = & \nonumber\\
& \rm \log_{10}(L^{\prime}_{CO(1-0)}[K\,Km\,s^{-1}\,pc^2]) - 4.31 & \label{eq:cfir} \\
& \rm \log_{10}[L_{[CII]}/L_{CO(1-0)}] = 3.6  &  \label{eq:cfir2}
\end{align}
Combining Equation~\ref{eq:cfir} and \ref{eq:cfir2}, we have the ratio of $\rm \log_{10}[L^{\prime}_{CO(1-0)}/L_{[CII]}] = 0.9$, which is used to convert the CO(1-0) LF in $\rm L^{\prime}_{CO(1-0)}$.  
The first relation is based on the equations defined in \S\,2.4 in \citet{Carilli2013}. 
The $\rm \log_{10}[L_{[C\,II]}/L_{CO(1-0)}]$ ratio  can span a range of $3 - 3.8$ as measured from a sample of normal galaxies at $z\sim0$ and $z\sim1-2$ \citep{Stacey1991,Stacey2010}. \citet{Zanella2018} recently examined a larger sample of galaxies with published $\rm L_{[CII]}$ and $\rm L_{CO}$ measurements at $z\sim 0 - 6$. We adopt the value of 3.6 in Equation~\ref{eq:cfir2}, consistent with the calculation of molecular mass described in \S\,\ref{sec:gasmass}.  Our adopted value makes a conservative conversion to [C\,II] luminosity. Extremely metal-poor, low mass, blue compact dwarf galaxies can have one order of magnitude higher [C\,II]-to-CO(1-0) ratio \citep{Cormier2014}. But these are unlikely to be good analogues, since the averaged star formation efficiency for the ALPINE sample is an order of magnitude higher than the local sample \citep{Faisst2019,Madden2013, Cormier2014}. In addition, our adopted value ignores that at $z\sim 4 - 6$, CMB photons will increase line excitation as well as background for CO(1-0) line \citep{daCunha2013}. The CO(1-0) line would be suppressed by $\sim$20\%\ at $z\sim6$ \citep{Vallini2018}. This would potentially push these estimated [C\,II] LFs to even higher luminosities, but is still well within the overall uncertainties in this conversion relation.

Figure\,\ref{fig:clf+fir+co} shows that within the large uncertainties, there is good agreement between all these constraints on the [C\,II] luminosity function at $z\sim4-6$. Combined, they strengthen even further our conclusion that  there has been strong evolution in the number densities of $L_{[CII]} > 10^{9}L_\odot$ sources between $z\sim5$ and $z\sim0$.

Finally, including the data from our work, Loiacono et al. (2020), the converted values using the FIR LFs (from ALPINE continuum data) from Gruppioni et al. (2020) and \citep{Koprowski2017}, we perform a fit to the data assuming a Schechter function. We derive $\rm log_{10}(L_{[CII]}^*/L_\odot) = 9.5\pm0.6$, $\phi^* = (8.4\pm8)\times10^{-4}$\,Mpc$^{-3}$\,dex$^{-1}$ and $\alpha=-1.1\pm0.3$\footnote{The fit was done using the {\sc curve$\_$fit} function within the {\sc scipy} package \citep{scipy}.}.  
The Schechter fit heavily relies on the FIR LFs and have large errors because both our and Loiacono et al results are only 
the lower and upper limits. It is expected that this Schechter form does not fit to either our data nor Loiacono et al.  
Performing the functional form fit here is for the convenience when making comparisons and when computing [C\,II] line luminosity density. At face value, the $z\sim 4 - 6$ $\rm L_{[CII]}^*$ has increased by a factor of 20 compared to $\rm L_{[CII]}^*\sim 2.2\times10^8L_{\odot}$ at $z\sim0$ \citep{Hemmati2017}. We caution that systematic errors could be large.

\subsection{Comparison with theoretical predictions} \label{sec:results_theory}

\begin{figure}
    \includegraphics[scale=0.45]{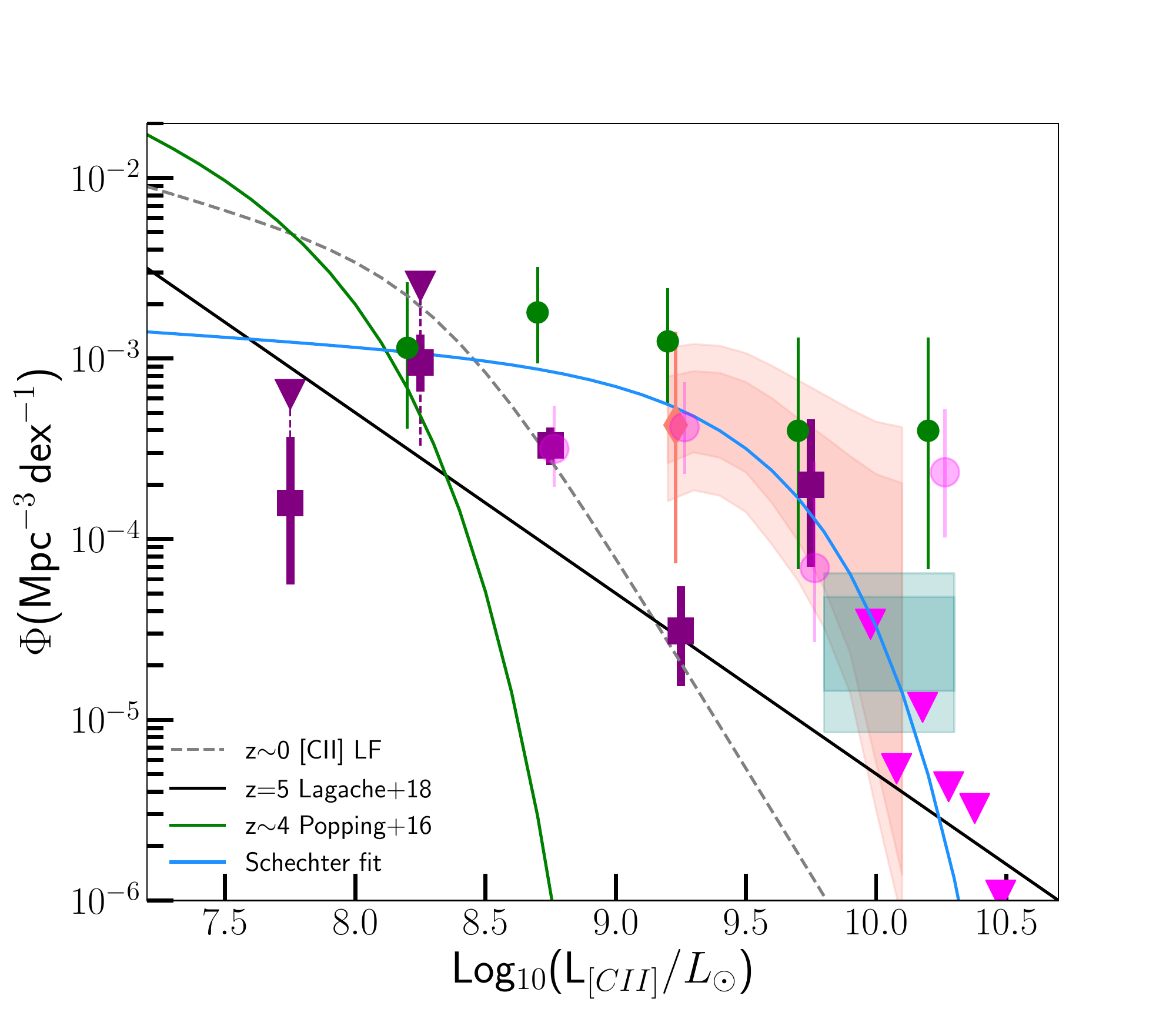}
    \caption{The comparison of the [C\,II] LF measurements from both the UV (purple) and [C\,II] serendipitous samples (green and pink dots) with the model predictions (black and green lines). Here we also included the [C\,II] LFs converted from the FIR and CO LFs. See text for the details.
    For additional figure legend see Figure~\ref{fig:clf+fir+co} for the data points.}
    \label{fig:clfmodels}
\end{figure}

Figure~\ref{fig:clfmodels} shows the model [C\,II] LFs from \citet{Popping2016} and \citet{Lagache2018} overlaid on the same data as shown in Figure\,\ref{fig:clf+fir+co}.
These models combine semi-analytic galaxy evolution models with radiative transfer calculations to make predictions on [C\,II] line emissions and LFs.  As already noted in our $z\sim0$ [C\,II] LF paper \citep{Hemmati2017}, the \citet{Popping2016} model is definitely ruled out by the observations. \citet{Lagache2018} model is much improved, but still fails to match the abundant luminous [C\,II] emitters at $z\sim 4 - 6$. The \citet{Lagache2018} model is reasonably consistent with the data at $\rm log_{10}(L_{[C\,II]}/L_{\odot})>10$, but the data overall are more consistent with a Schechter functional form than the power law form of this model. 

Several cosmological simulations have investigated CO(1-0) and [C\,II] emission at high-$z$, most of them focusing on the L([CII])-SFR relation and associated deviations. As illustrated in \citet{Ferrara2019}, the physical processes of [C\,II] emission are complex, and the assumed gas density in PDRs (or size and filling factor), ionization factor, metallicity and starburstness can all affect [C\,II] luminosity. Some models can reproduce both the luminous and weak [C\,II] sources in the L([CII])-SFR plane \citep{Vallini2015, Lagache2018, Ferrara2019}. The discrepancy between our LF and the models however implies that the current models for synthesizing the high-$z$ [C\,II] populations likely involves other factors that are still not well understood. More work is clearly needed.

\subsection{Implications to the molecular gas mass density at $z\sim4 - 6$} \label{sec:gasmass}
It has been widely recognized that the evolution of the molecular gas mass as a function of redshift holds clues to understanding galaxy formation and evolution more broadly \citep[for a review see][]{Carilli2013}. In particular, we want to compare with the model for $\rho_{molgas}$ as a function of $z$ from \citet{Liu2019}. 
\citet{Liu2019} combined their ALMA galaxy catalog with 20 other published datasets and made a comprehensive analysis of scaling relations between stellar mass, gas fraction and redshift. 
They expressed the molecular gas mass density, $\rho_{mol}(z) = \sum_{M_*} \Phi_{SMF}(z,M_*)\times M_* \times \mu_{gas}(z,M_*,\Delta MS)$, as a function of stellar mass function $\Phi$ and gas fraction $\mu$. They integrate down to $M_*=10^9$M$_{\odot}$.  Using the \citet{Liu2019} $\mu_{molgas}=M_{molgas}/M_*\sim2$ at $z\sim5$ and $\alpha_{[CII]-M_{molgas}} \sim 30M_{\odot}$/L$_{\odot}$ from \citet{Zanella2018}, this lower limit translates to $\rm L_{[CII]}\sim7\times10^7$L$_{\odot}$. Therefore we start by integrating the [CII] luminosity function down to this limit giving us a luminosity density of $\mathscr{L}_{[CII]}$ is $1.2\times10^6 L_\odot$\,Mpc$^{-3}$. \citet{Zanella2018} find that $\alpha_{[CII]-M_{molgas}}$ relation has an $1\sigma$ of 0.3 dex, translating into a linear ratio $\rm M_{mol}/L_{[CII]} = 15 - 60\,M_\odot/L_\odot$. We calculate the cosmic volume averaged molecular mass density $\rho_{molgas}=(2 - 7)\times10^7M_\odot$/Mpc$^3$ at $z\sim 4 - 6$. At $2\sigma$ and 95\%\ confidence level, we estimate $\rho_{molgas}=(1 - 14)\times10^7M_\odot$/Mpc$^3$. 

Note however, that when using the \citet{Zanella2018} relation on individual ALPINE galaxies to convert their $L_{[CII]}$ into the molecular gas mass, we see that the mean gas fraction is $\mu_{molgas}\approx3$ \citep{Dessauges2020}. This is slightly higher than in the one of the \citet{Liu2019} sample likely due to the fact that the ALPINE galaxies have lower stellar masses than the \citet{Liu2019} sample. We checked that changing this constant does not change our results in any substantive way (see the discussion below).

\begin{figure}
    \includegraphics[scale=0.38]{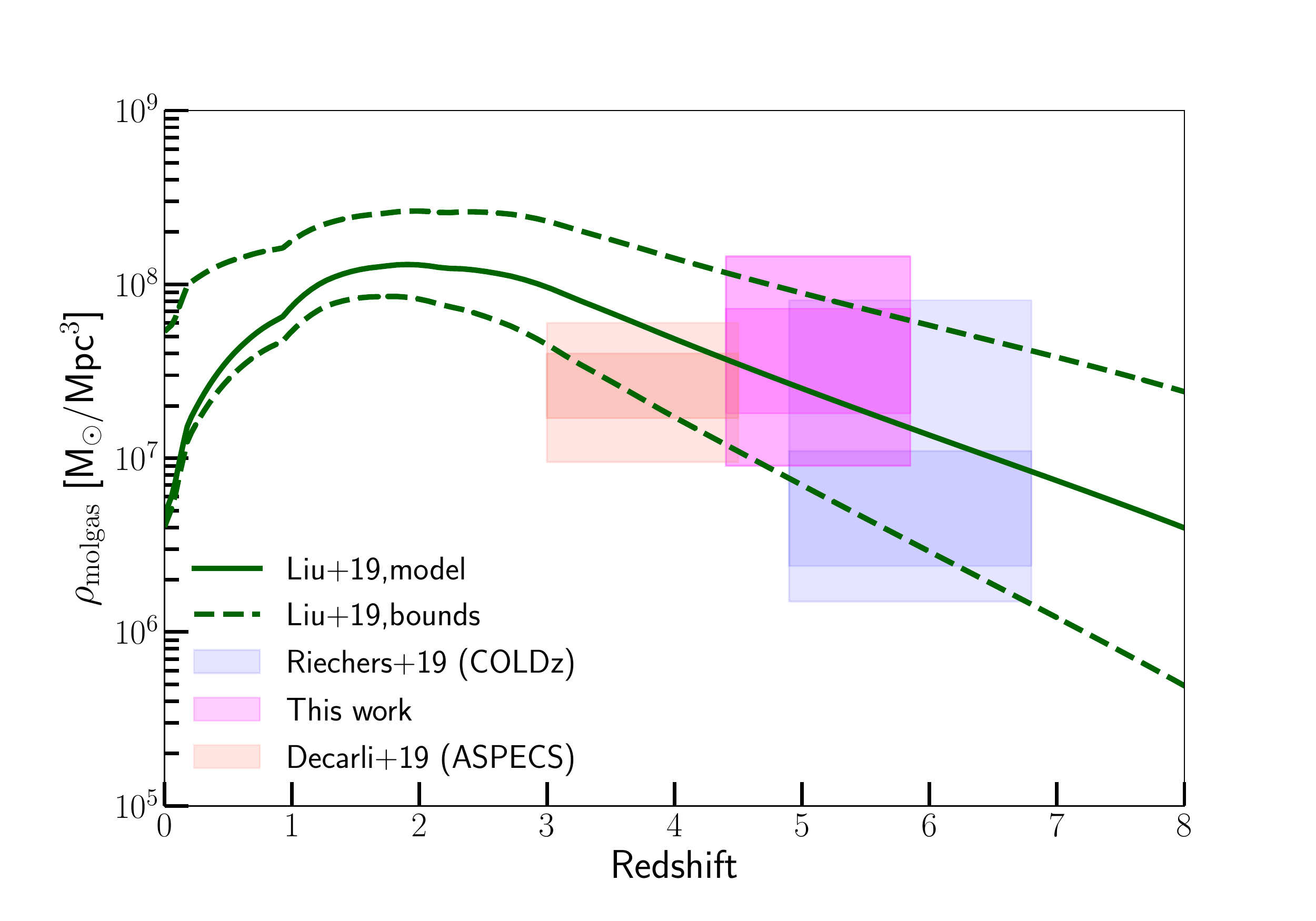}
    \caption{The estimated mass density of molecular gas based on our [C\,II] LF (pink shaded region). The height of this pink region corresponds to $2\sigma$ and $1\sigma$ of the M$_{mol}$-L$_{[CII]}$ scaling constant. Similarly, the ASPECS and COLDz results (salmon and light green shaded regions) also include $2\sigma$ and $1\sigma$ uncertainties.}
    \label{fig:molmass}
\end{figure}

In Figure\,\ref{fig:molmass}, we overlay our estimates of $\rho_{molgas}$ on the \citet{Liu2019} model predictions (based on the A3COSMOS sample). We also overlay the 1 and 2$\sigma$ constraints from ASPECS \citep{Decarli2019} and COLDz \citep{Riechers2019} which surveys measured CO transitions using ALMA and JVLA and set constraints on the mass density of molecular gas at $z\sim 3 - 6$.  This cross-check is critical because all these surveys have pros and cons in their attempts to constrain the evolution of the cosmic molecular gas density. The key advantage of the ALPINE survey is its large sample (thanks to the [C\,II] line being a significantly brighter tracer of molecular gas than any CO line), but due to the uncertainties in the selection function we are left with only lower and upper bounds on the LF. The A3COSMOS-based constraint is also affected by selection biases. Both ASPECS and COLDz, while smaller samples, have the key advantage of being unbiased in terms of cold gas selection, since they use CO which is the most reliable tracer of cold molecular gas mass currently known. Figure\,\ref{fig:molmass} shows that, within the uncertainties, all these studies are consistent. As such, these surveys provide
supporting evidence that we are obtaining useful constraints as we use the sizeable ALPINE sample to constrain the cold gas density vs redshift relation. 

Figure\,\ref{fig:molmass} shows that our results are well within the range of the \citet{Liu2019} model, only slightly favoring the upper half. The spread in the model shown indicates different assumption of the molecular gas fraction as a function of redshift. The upper limit is based on \citet{Scoville2017} analyses of the ALMA dust continuum observations at $z<3.8$, and the lower limit is based on the \citet{Tacconi2018} relation, primarily from the PHIBBS CO survey at $z<2.5$. In contrast to our results, the results of the CO surveys (COLDz and ASPECS) favor the lower half of the models shown. However, given the large uncertainties in all of these estimates, this slight tension between us and the CO survey is not too significant. Within the 2$\sigma$ uncertainties, our results are consistent with those from COLDz and ASPECS. 

\citet{Dessauges2020} also examined the molecular gas content of individual ALPINE galaxies at z~5, using [C\,II] detections to constrain the total molecular gas mass at $z\sim5$. Like us, they use the scaling relations from \citet{Zanella2018}. They find $\mu_{molgas}\approx3$ which leads to slightly lower total mass density and a better agreement with the \citet{Tacconi2018} model (see Fig. 7 of their paper). Indeed, given the large uncertainties on these molecular gas mass densities, we cannot really exclude any of the models. The main difference between our analyses however is that our calculation is based on our best estimate of the [CII] luminosity function, including all three ALPINE constraints thereof (the UV-sample, the serendipitous sample and the continuum-sample) as well as the far-IR LF from \citet{Koprowski2017}. In this paper we do not look at the ALPINE constraints on the star-formation rate density at $z\sim5$. This question is examined in our companion paper, Loiacono et al. (2020).

\section{Summary and conclusions}
\label{sec:summary}

The ALPINE survey carried out ALMA band-7 observations of 118 galaxies at $z \sim 4.40 - 5.85$ selected by their UV luminosity ($M_{1500\angstrom} < -20.2$, or $L_{1500\angstrom} > 0.6L^*_{1500\angstrom}$) within the COSMOS and GOODS-South fields.  
The observations show a high fraction (75/118) of [C\,II] detections confirming that, at high-$z$, [C\,II] line emission broadly traces the UV emission, especially for galaxies without substantial dust obscuration. 

Using the ALPINE UV selected target sample, we calculate the lower bounds to the true volume densities of [C\,II] emitters as a function of line luminosity. 
Our [C\,II] LFs are consistent with that of $z\sim0$. The possible higher volume density of [C\,II] emitters at $\rm L_{[CII]} > 10^{9.5}L_\odot$ is only a $(1-2)\sigma$ result (68\%\ - 95\%\ confidence level), thus requires further confirmation. 
We compare our lower limits results with the upper bounds on the [C\,II] LFs derived from the serendipitous [C\,II] sample (Loiacono et al. 2020), and with the inferred [C\,II] LFs converted from the FIR and CO LFs at $z\sim4-6$. The differences suggest that there are potentially more UV-faint and [C\,II] bright galaxies missing from the ALPINE sample. This type of sources could push the true [C\,II] volume density higher than that of $z\sim0$. This speculation requires future confirmation.

We fit a single Schechter function model to both these ALPINE constraints as well indirect [C\,II] LFs converted from published far-IR and CO(1-0) LFs at $z\sim4-5$. This model fit suggests that the population of luminous [C\,II] emitters at $z\sim4-6$ is roughly $20-50$\,$\times$ more abundant than that of the local Universe.

By integrating the [C\,II] LF, and adopting scaling relations between [CII] uminosity and molecular gas mass we estimate that the molecular mass density at $z\sim 4 - 6$ reaches roughly $(2 - 7)\times 10^7M_\odot$\,Mpc$^{-3}$ ($1\sigma$). This value is consistent with the evolutionary tracks of $\rho_{mol}(z)$ from \citet{Liu2019}, favoring slightly models with somewhat higher galaxy gas fractions. By contrast, CO surveys such as COLDz and ASPECS \citet{Riechers2019,Decarli2019} favor slightly lower gas fractions. However, this difference is not significant at present given the large uncertainties in all the measurements.

Our measurements at $z\sim 4 - 6$ have important implications for upcoming millimeter spectral surveys such as CONCERTO \citep{Lagache2018b} and CCAT-Prime \citep{Stacey2018}. They provide benchmarks of the expected number of sources, in particular predicting that large area millimeter spectral surveys should find many extremely luminous [C\,II] emitters at $z\sim 4 - 6$. 

Finally, our results are in tension with the available model predicted LFs, particularly at the luminous end \citep{Popping2016, Lagache2018} suggesting model assumptions about the physical conditions in PDR at these redshifts may not be correct.  
Better understanding of the physical conditions in the ISM at these redshifts is clearly needed. This would be enabled by a far-IR spectroscopy survey facility such as the proposed Origins Space Telescope\footnote{\url{https://asd.gsfc.nasa.gov/firs/}}.

\acknowledgements
We thank Liu, Daizhong from Max-Planck-Institut f{\"u}r Astronomie for sending us their results in electronic form.
 This paper is based on
data obtained with the ALMA Observatory, under Large Program 2017.1.00428.L. ALMA is a partnership of ESO (representing
its member states), NSF(USA) and NINS (Japan), together with
NRC (Canada), MOST and ASIAA (Taiwan), and KASI (Republic of Korea), in cooperation with the Republic of Chile. The Joint
ALMA Observatory is operated by ESO, AUI/NRAO and NAOJ. GL acknowledges support from the European Research Council (ERC) under the European Union’s Horizon 2020 research and innovation programme (project CONCERTO, grant agreement No 788212) and from the Excellence Initiative of Aix-Marseille University-A*Midex, a French “Investissements d’Avenir” programme.
AC, CG, FL, FP and MT acknowledge the support from grant PRIN
MIUR 2017 - 20173ML3WW\_001. ST acknowledge support from the European Research Council (ERC) Consolidator Grant funding scheme (project “ConTExt”, grant number: 648179). The Cosmic Dawn Center (DAWN) is funded by the Danish National Research Foundation under grant No. 140.
Our research made use of Astropy,\footnote{http://www.astropy.org} a community-developed core Python package for Astronomy \citep{astropy:2013, astropy:2018}.
LV acknowledges funding from the European Union’s Horizon 2020 research and innovation program under the Marie Sklodowska-Curie Grant agreement No. 746119.
G.C.J. acknowledges ERC Advanced Grant 695671 ``QUENCH'' and support by
the Science and Technology Facilities Council (STFC). E.I.\ acknowledges partial support from FONDECYT through grant N$^\circ$\,1171710. JDS was supported by JSPS KAKENHI Grant Number JP18H04346, and the World Premier International Research Center Initiative (WPI Initiative), MEXT, Japan. This paper is dedicated to the memory of Olivier Le F\`evre, PI of the
ALPINE survey.

\bibliography{references}

\begin{deluxetable}{ccccc}
\tablecaption{ALPINE [C\,II] LFs at $z$\,$\sim$\,$4 - 6$ in $10^{-4}$\,Mpc$^{-3}$\,dex$^{-1}$ \label{tab:clf}}
\tablewidth{0pt}
\tabletypesize{\scriptsize}
\tablehead{
\colhead{Log$_{10}(L_{[CII]})$} &
\colhead{$\phi_{[CII]}^a$} &
\colhead{$\phi_{[CII]}^{up,b}$} &
\colhead{$\phi_{[CII]}$} &
\colhead{$\phi_{[CII]}^{up,c}$} \\ 
\hline\hline
\colhead{$L_\odot$} &
\multicolumn2c{$z\sim4.5$ } &
\multicolumn2c{$z\sim5.5$} 
}
\startdata
7.75 & ...           & ...          & $1.58^{+2.09}_{-1.02}$  & $11.1^{+2.09}_{-1.02}$  \\
8.25 & $7.98^{+3.94}_{-2.77}$ & $26.4^{+3.94}_{-2.77}$  & $11.1^{+4.20}_{-3.15}$  & $24.0^{+4.20}_{-3.15}$ \\ 
8.75 & $4.12\pm0.76$ & ...          & $2.47^{+0.94}_{-0.70}$  & ...          \\ 
9.25 & $0.55^{+0.33}_{-0.22}$ & ...          & $0.074^{+0.072}_{-0.04}$ & ...        \\ 
9.75 & $2.0^{+2.63}_{-1.29}$  & ...          & ...            & ...          \\
\hline\hline \\
7.95 & $0.59^{+0.78}_{-0.38}$ & ...  & $3.59^{+1.93}_{-1.32}$ & ... \\ 
8.45  & $9.72^{+2.30}_{-2.30}$ & ... & $10.8^{+3.92}_{-2.97}$ & ... \\ 
8.95  & $2.31^{+0.49}_{-0.49}$ & ... & $0.76^{+0.37}_{-0.26}$ & ... \\
9.45 & $2.03^{+1.61}_{-0.97}$ & ...  & ...  & ... \\ 
\enddata
\tablecomments{{\it \bf a}: $\phi_{UV}$ is the [C\,II] LF derived based on the ALPINE primary targets, {\it i.e.} the UV sample. {\it \bf b \&\ c}: $\phi_{UV}^{up}$ refers to the corrected [C\,II] LFs by taking into account of the [C\,II] non-detections in the ALPINE sample. These mark the upper limits to the two lower luminosity bins because we assume that these non-detections have $3\sigma$ line luminosity. 
}

\end{deluxetable}


\end{document}